\documentstyle [12pt,epsf]{article}

\newcommand{\beq}{\begin{equation}}
\newcommand{\dd}{\partial}
\newcommand{\eeq}{\end{equation}}
\newcommand{\bea}{\begin{eqnarray}}
\newcommand{\eea}{\end{eqnarray}}

\newcommand{\vf}{\varphi}

\newcommand{\e}{{\cal E}_\omega}
\begin{document}
\baselineskip 7.2 mm

\def\thefootnote{\fnsymbol{footnote}}

\begin{flushright}
\begin{tabular}{l}
CERN-TH/97-259 \\
\end{tabular}
\end{flushright}

\vspace{12mm}

\begin{center}

{\Large \bf

Supersymmetric Q-balls as dark matter
}
\vspace{18mm}

\setcounter{footnote}{0}

Alexander Kusenko\footnote{ email address: Alexander.Kusenko@cern.ch}
and
\setcounter{footnote}{6}
Mikhail Shaposhnikov\footnote{ email address: mshaposh@nxth04.cern.ch}
\\
Theory Division, CERN, CH-1211 Geneva 23, Switzerland \\

\vspace{36mm}

{\bf Abstract}
\end{center}
 Supersymmetric extensions of the standard model generically contain
stable non-topological solitons, Q-balls, which carry baryon or lepton
number. We show that large Q-balls can be copiously produced in the
early universe, can survive until the present time, and can contribute
to dark matter.

\vspace{30mm}

\begin{flushleft}
\begin{tabular}{l}
CERN-TH/97-259 \\
September, 1997
\end{tabular}
\end{flushleft}

\vfill

\pagestyle{empty}

\pagebreak

\pagestyle{plain}
\pagenumbering{arabic}
\renewcommand{\thefootnote}{\arabic{footnote}}
\setcounter{footnote}{0}

\pagestyle{plain}

\section*{Introduction}

The nature of the missing matter in the universe remains one of the
most intriguing outstanding problems in particle physics and cosmology.
Stable particles hypothesized in theories beyond the standard model
have been considered as candidates for dark matter. In the
supersymmetric extensions of the standard model, the lightest
supersymmetric particle (LSP), whose stability may be guaranteed by the
conservation of R-parity, can naturally contribute to the hidden mass.

Recently, it has been pointed out that non-topological solitons,
Q-balls \cite{lp,coleman}, are generically present in the MSSM
\cite{ak_mssm}. Q-ball is a coherent state of a complex scalar field,
whose existence and stability are due to the conservation of some
global U(1) quantum number. In the MSSM the usual baryon and lepton
numbers may play the role of such conserved quantity for the Q-balls
built of squarks and sleptons, respectively. One can ask, whether these
objects could have formed in the early universe, and whether their
lifetime can be long enough for them to survive until present and
contribute to dark matter.

Several mechanisms could have lead to the formation of Q-balls in the
early universe. Non-topological solitons can be created \cite{s_gen} in
the course of a phase transition (``solitogenesis''), or they can be
produced via fusion \cite{foga,gk,ak_pt} in a process reminiscent of
the big bang nucleosynthesis (``solitosynthesis''). Finally, small
Q-balls \cite{ak_qb} can be pair-produced at high temperature. None of
these scenarios, however, seems to be capable of producing large enough
Q-balls to survive the subsequent evaporation \cite{evap} in the
presence of massless fermions that carry the same global charge. One
might think, therefore, that, in the absence of a viable mechanism that
could naturally produce very large solitons, no relic Q-balls could
remain in the present universe.

In this paper, we show that there exists an effective mechanism for the
production of Q-balls with enormous charges, big enough to survive
until present. The basic idea is the following. A complex scalar field
$\vf$ inside the Q-ball is in a coherent state with time-dependent
phase: $\vf=R(x) e^{i \omega t}$, where the radial component $R(r)$,
$r=\sqrt{\vec{x}^2}$, tends to zero as $r \rightarrow \infty$. An
infinite size Q-ball with $R(r) = const$, Q-matter \cite{coleman}, is
similar in nature to the coherent states of a scalar condensate often
encountered in inflationary cosmology. For example, formation of the
scalar condensate with non-zero baryon number is the starting point for
the Affleck-Dine scenario for baryogenesis \cite{ad}. In this scenario,
a combination of squarks and sleptons, or some other fields carrying a
baryon or lepton number, has a large expectation value along some flat
direction of the potential at the end of inflation. At large VEV, the
baryon number can be strongly violated by the high-scale physics. As a
result of the baryon number non-conservation, along with the CP
violation, the scalar condensate acquires a baryon number. The
subsequent evolution leads it into the domain of conserved baryon
number. From this point on, it can be thought of as Q-matter.

Depending on the dynamics, an initially spatially-homogeneous scalar
condensate that carries a conserved U(1) charge may become unstable
with respect to small coordinate-dependent perturbations and develop a
spatial pattern that comprises domains of high and low charge density.
If a theory admits non-topological solitons \cite{lp,coleman}, the
lowest energy state in a given charge sector is a Q-ball. Very large
(and, thus, stable) Q-balls can, in fact, be produced this way.

A related issue is that of the string moduli. If their evolution leads
to formation of isolated solitons, it can help alleviate the
cosmological moduli problem \cite{moduli,gia}. The weakly-interacting
moduli may possess some energy density during or after the
nucleosynthesis. However, if this energy density is incarcerated inside
Q-balls, it may be harmless.

In what follows, we first discuss the stability and decay of Q-balls.
Then we analyze formation of Q-balls through solitosynthesis and
collisions. We then consider the evolution of a complex scalar field
and describe the formation of a spatial pattern that can evolve into
Q-balls. Finally, we discuss the contribution of relic Q-balls to dark
matter and comment on the possibility of solving the cosmological
moduli problem through the incarceration of malicious moduli.

\section{Lifetime of Q-balls}

The fate of Q-balls in the early universe is ultimately determined by
their lifetime. If the soliton in question is made of a scalar that has
no interactions with light fermions, then it is entirely stable. If
such objects had formed through the breakdown of a homogeneous
condensate, they would have survived until the present and would
contribute to the matter density of the universe as a form of dark
matter.

If, however, the scalar particles can decay into lighter fermions that
carry the same global charges, as is the case with the squarks and
sleptons of the MSSM, then it is the evaporation rate that determines
the lifetime of a Q-ball \cite{evap}. Clearly, it is only the fermionic
decay modes that ought to be considered in this respect, because, in
the sector of scalar degrees of freedom, the Q-ball is the state of
minimal energy, and it cannot decay into bosons.

Depending on the scalar potential $U(\vf_1,...,\vf_n)$, the mass of a
large Q-ball may grow with charge as $m_{_Q} \sim Q^p$, where $0<p<1$.
If the quantity

\beq
\frac{U(\vf_1,...,\vf_n)}{\sum q_i \vf_i^2},
\label{Uoverphi2}
\eeq
where $q_i$ is the charge of the field $\vf_i$ (some of the charges may
be zero), is minimized at a finite value of $\vf$, then the mass of a
large Q-ball is proportional to the first power of $Q$ \cite{coleman}.
In this case the mass per unit charge does not depend on the size of
the Q-ball.

However, if $U(\vf)$ is essentially flat, {\it i.\,e.} if it grows
slower than the second power of $\vf$, then, in general, $p<1$
\cite{dks}. Such potentials can arise naturally in theories with
supersymmetry breaking communicated at low energy, for example, by
gauge interaction (for review, see, {\it e.\,g.}, Ref. \cite{gauge} and
references therein). If the potential is essentially flat for large
$\vf \gg m_{_S}$, then $m_{_Q} \sim Q^{3/4}$ \cite{dks}. We
recapitulate briefly the argument of Ref. \cite{dks}. The field inside
the Q-ball is $\vf_i(x,t)= \vf_i(x) \exp \{ i q_i \omega t\}$
\cite{ak_mssm}. The value of $\omega$ is determined by extremizing the
functional

\beq
{\cal E}_\omega = \int d^3x \, \left [\frac{1}{2} \sum_k |\nabla \vf_k
|^2
+ \hat{U}_\omega(\vf)
\right ] + \omega Q ,
\label{Ewt}
\eeq
where $\hat{U}_\omega (\vf) = U(\vf)\ - \ \frac{1}{2} \omega^2 \,
\sum_k q_k^2 \, |\vf_k|^2$. Written in terms of dimensionless variables
$\xi = \omega x$ and $\psi_i = \vf_i/\omega $, $\e \approx a \omega +
b/\omega^3 + \omega Q$, if $U(\vf) \approx const$. Here $a$ and $b$ are
independent of $\omega$. Therefore, the extremum of functional
(\ref{Ewt}) is achieved for $\omega \propto Q^{1/4}$, which corresponds
to

\beq
m_{_Q} \propto Q^{3/4}.
\label{E_Q3_4}
\eeq

By virtue of the relation (\ref{E_Q3_4}), a sufficiently large Q-ball
is entirely stable against decay into all but massless fermions because
its energy per unit charge can be less than the mass of the lightest
fermion. In the MSSM, there are Q-balls (B-balls) made of squarks and
Higgs fields \cite{ak_mssm}, for which the baryon number plays the role
of the U(1) charge discussed above. The lightest baryons, nucleons,
have mass of order $m_n \sim 1$ GeV. A B-ball is entirely stable if it
has charge

\beq
Q \equiv B \stackrel{>}{_{\scriptstyle \sim}} \left( \frac{m_\vf}{m_n}
\right )^4 \ \sim \ 10^8,
\label{stableQ}
\eeq
where we took $m_\vf\sim 10^2$ GeV for the squark mass.

If the Q-ball has a lepton charge, but no baryon charge, then it can
evaporate \cite{evap} by emitting lepton number from the surface with
(nearly massless) neutrinos. It is beyond the scope of the present work
to evaluate the decay width of slepton Q-balls in the MSSM taking into
account the variety of fermionic decay modes. Instead, we will obtain a
rough estimate by adopting the discussion of Ref. \cite{evap} to our
case.

The model of Ref. \cite{evap} had a Yukawa coupling of a scalar field
$\vf$ to massless fermions. Inside the Q-ball, the Dirac sea of
fermions is filled; and the Fermi pressure inhibits further decay. The
decay of the scalar condensate can, therefore, proceed through the
surface only, and its rate is suppressed by the surface-to-volume ratio
for large Q-balls. The rate of such decay is \cite{evap}

\beq
\frac{d Q}{dA\,  dt} < \frac{\omega^3}{192 \pi^2} ,
\label{decay}
\eeq
where $A$ is a Q-ball surface area. For the lifetime of an L-ball to  
exceed the age of the Universe, its charge should be $Q>10^{32}$.

The effects of gravity on Q-balls are very small and will be neglected,
as long as the size of the soliton is much larger than the
corresponding Schwartzschild radius. For a Q-ball in a flat potential,
this is the case if $Q\ll Q_g= (m_{_P}/m_{_\vf})^4 \sim 10^{64}$ for
$m_{_\vf} \sim 100$ GeV. A Q-ball with charge that exceeds $Q_g$ would
be a black hole. It would evaporate through Hawking radiation, and its
global charge would be lost. In our analyses we never encounter Q-balls
with such large charges, and we don't know of any mechanism by means of
which they could form.

If the potential is not flat, then $E_{_Q} \sim Q$ \cite{lp,coleman},
and the Q-ball would decay through evaporation \cite{evap} regardless
of the value of its charge. (The charge needed to survive the
evaporation would exceed the black hole limit by many orders of
magnitude.)

At finite temperature, the existence of Q-balls depends on the form of
the temperature-dependent effective potential. If it admits Q-balls,
then the thermal processes do not affect an existing soliton. If,
however, a Q-ball is formed out of equilibrium at some temperature $T$  
in a
 system that does not allow Q-ball solutions, the
collisions with particles in the hot plasma can lead to an erosion of
the condensate. For large solitons this process is strongly suppressed
by the small penetration depth of the outside particles inside the
Q-ball. A particle $\psi$ that interacts with the field $\vf$ receives
a contribution to its mass of order $m_\psi(x) \sim g \langle \vf (x)
\rangle$ inside the Q-ball; outside Q-ball $m_\psi \sim g T$. At
temperature $T$ a $\psi$ particle cannot penetrate inside the Q-ball
beyond the point $x$ where $\vf(x) \sim T/g$. The condensate can be
thought of as a coherent state of $\vf$ particles with mass $m_{cond}
\sim m_\vf/Q^{1/4}$ and density $\omega \vf^2(x)$. If a Q-ball is
created at the time $t_i$, the total number of collisions with the
$\psi$ particles in plasma during the lifetime of the cooling universe
is
\beq
\int_{r_0}^{\infty} r^2 dr \, \int_{t_i}^{\infty} dt \ T^3(t) \, \sigma
\omega \, \vf^2(r) \, v_\psi(r),
\label{collisions}
\eeq
where $\sigma \sim g^2/T^2$ is a cross-section of particle-particle
collision, the velocity of the $\psi$ particle at a given point is
determined by energy conservation. The stopping point, $r_0$ is defined
by the relation $m_\psi(r) = T$, or $g \vf (x) = T$. For squark B-balls
$g\sim 1$. One can use $T^2=M_0/t$ to relate temperature to time in a 

radiation dominated universe. Here and below $M_0=M_{P}/(1.66 \,
n_{eff}^{1/2})$, and $n_{eff}$ is the number of effective degrees of
freedom. In each collision, only a fraction of energy, $\delta E \sim T
(m_{cond}/m_\psi)^2$, is transmitted to the condensate. Therefore,
using (\ref{collisions}), the total energy transmitted to the Q-ball is

\beq
\Delta E \sim R^2 \ m_{cond}\  \omega \ M_0 \ T^2 \sigma
\int_{r_0}^{\infty}
\frac{\vf^2(r) v(r)}{m_\psi(r)}  dr .
\eeq
This is sufficient to knock out $\Delta Q \sim \Delta E/m_\vf$
particles from the condensate. For a ``flat'' potential,

\beq
\Delta Q \sim \left(\frac{T}{m_{\vf}}\right)^2 
\left ( \frac{M_0}{m_{\vf}} \right ).
\label{erosion}
\eeq
A Q-ball can survive the erosion by thermal plasma if $\Delta Q < Q$.

\section{Solitosynthesis and collisions}

Solitosynthesis of non-topological solitons \cite{foga,gk,ak_pt} is the
process of charge accretion that may take place in a system at finite
temperature with non-zero charge asymmetry $\eta$. Q-ball is the
minimum of energy $E$ in the sector of fixed charge, but is also a
state of small entropy $S$ (because it is an extended object). For high
temperature it may not be the minimum of free energy $F=E-TS$ because
of the $TS$ term. At some small enough temperature $T_{0}\sim
m_\vf/|\ln \eta |$, however, the second term in $F$ is no longer
important, and the gain from minimizing $E$ overwhelms the loss from
lowering the entropy \cite{foga,gk,ak_pt}. At that point, Q-ball is the
minimum of free energy. In the absence of light charged fermions, a
copious production of Q-balls can take place.

However, if there are light fermions carrying the same global charge as
the scalars, then the charge asymmetry in the state of the lowest free
energy is accommodated entirely in the fermionic sector, thus making
solitosynthesis impossible. For example, the solitosynthesis of
baryonic and leptonic Q-balls in the MSSM is only possible in a B or L
breaking minimum of the potential, where the quarks or leptons are
sufficiently massive \cite{ak_pt}.

Even in the absence of light charged fermions, solitosynthesis is
unlikely to produce large Q-balls unless the charge asymmetry is very
large \cite{foga,gk}. The baryon asymmetry of the universe is as small
as $10^{-10}$, which is certainly not sufficient for synthesizing large
enough B-balls to satisfy the stability bound (\ref{stableQ}). One can
ask whether the subsequent mergers of small Q-balls can increase the
average size considerably.

Regardless of the mechanism that lead to the formation of Q-balls,
their size can further change because of the collisions in which two
Q-balls would merge and form a larger soliton. For simplicity we will
neglect the evaporation of Q-balls assuming that they are made of
stable scalars. The effect of collisions is described by the kinetic
equation

\beq
\frac{\dd}{\dd t} N_{_Q} + 3 H N_{_Q} =
\left [ \frac{1}{2} \int \langle \sigma_{_Q} v \rangle
N_{Q-q} N_q \, dq - N_{_Q} \int \langle \sigma_{_Q} v \rangle
N_q \, dq \right ],
\label{clsn1}
\eeq
where $N_{_Q}$ is the density of Q-balls with charge $Q$, and $H$ is
the Hubble constant. We note in passing that if the individual
particles are included as $Q=1$ objects, the same equation
(\ref{clsn1}) can describe the charge accretion via absorption of
particles by Q-balls. The cross-section for the merger of two Q-balls
is determined mainly by their geometrical size, $\sigma_{_Q} \sim 4 \pi
R_{_Q}^2$, and the average velocity of their non-relativistic Brownian
motion is $v\sim \sqrt{ T/m_{_Q}}$. For a thick-wall Q-ball in a flat
potential the quantity $\langle \sigma_{_Q} v \rangle \propto
R_{_Q}^2/\sqrt{m_{_Q}} $ changes with charge very slowly, as $Q^{1/8}$.
Therefore, for a large range of charges, it can be considered constant.
One can define the reduced number density of Q-balls as $n_{_Q}=N_{_Q}
/ T^3 $. Then
\beq
\frac{\dd}{\dd \tau} n_{_Q}=
\int n_{Q} n_{Q-q} dq - 2 n_Q \int n_q dq,
\label{clsn2}
\eeq
where $\tau= \tau_{max} [1-(T/T_0)^{3/2}] $ and varies from $\tau=0$ to
$\tau_{max}= M_0 \langle \sigma_{_Q} \rangle T^{3/2}/\sqrt{m_{_Q}} $.
This equation can be solved by applying a Laplace transform to
disentangle the integrals on the right-hand side. The resulting
equation is a first-order differential equation for $f_s(t) \equiv
\int_0^\infty n_{_Q} \exp (-Q s) dQ$,

\beq
\frac{\dd}{\dd \tau} f_s=
 f_s^2- 2 f_s f_0 .
\eeq

Let us consider, for example, an initial distribution that peaks near
$Q=Q_{0}$. The value of $Q_0$ can be set by some earlier process, for
example, solitosynthesis. We require that at $\tau=0$, $n_Q(1)
=\epsilon \delta(Q-Q_0)$, and, hence, $f_s(1)= \epsilon \exp \{-Q_0
s\}$. This initial condition corresponds to a set of equal size Q-balls
with charge $Q_0$ each and number density $\epsilon$ that start merging
through collisions at $\tau=0$. The solution is
\beq
f_s(\tau)=\frac{\epsilon}{(1 +\epsilon \tau)^2} \frac{1}{e^{Q_0 s}-
\frac{\epsilon \tau}{1+ \epsilon \tau}}.
\label{sol_s}
\eeq
The inverse Laplace transform of $f_s(\tau)$ in equation (\ref{sol_s})
is a contour integral that can be found by summing over the residues of
the function $(f_s(\tau) \exp \{Q s\})$ in the complex plane.

\bea
n_{_Q}& = &
\frac{\epsilon}{(1 +\epsilon \tau)^2}
\left [\frac{\epsilon \tau}{1 +\epsilon \tau }
\right ]^{(Q-Q_0)/Q_0}
\sum_n \delta(Q/Q_0-n),
\label{distr}
\eea
where $n$ is an integer. The distribution (\ref{distr}) describes an
array of delta-functions at integer values of $Q/Q_0$ weighted
differently for different $\tau$. This is to be expected because each
collision changes the charge of a Q-ball by an integer multiple of
$Q_0$.

As temperature goes to zero, $\tau \rightarrow \tau_{max}$, the
distribution of Q-balls approaches (for $\epsilon \tau_{max} \gg 1$)

\beq
\lim_{\tau \rightarrow \tau_{max}} n_{Q}(t) \propto
\frac{\epsilon}{(\epsilon \tau_{max})^2}
\exp
\left ( - \frac{1}{\epsilon \tau_{max}}
\frac{Q}{Q_0}
\right )
\eeq
for integer $Q/Q_0$. This distribution is suppressed for large
values of $Q/Q_0 \gg \epsilon \tau_{max} $.

This process, which, for reasons explained earlier, is only relevant
for weakly interacting stable particles, may considerably increase the
average size of a Q-ball.

\section{Instability and pattern formation in the motion of a scalar
  condensate}

We now turn to the mechanism that can, in fact, lead to the formation
of very large Q-balls in the early universe. In particular, huge
Q-balls made of squarks and sleptons can be produced this way.

Evolution of a spatially homogeneous condensate carrying a baryon or
lepton number has been studied in detail in connection with the
Affleck-Dine \cite{ad} mechanism for baryogenesis. The scalar under
consideration is a complex field that carries some U(1) charge ({\it e.
g.}, the baryon number) and can be a combination of squarks and
sleptons of the MSSM \cite{drt}. If the U(1) charge is preserved by the
low-energy physics, the eventual decay of such condensate can
contribute to the corresponding charge asymmetry.

It is possible, however, that the spatially homogeneous condensate
develops an instability that leads to the formation of domains with
higher and lower charge density. If the theory admits Q-balls, such
pattern can evolve into Q-balls before the condensate is eroded by
decay. Thermal fluctuations cannot wash out the coherent state as long
as the scalar VEV is larger than $g T$, where $g$ is a typical
coupling. This is because all the particles coupled to $\vf$ with a
coupling $g$ have masses of order $g \langle \vf \rangle \gg T$. After
the Q-balls form, the subsequent decay of the condensate may become
impossible (if the Q-ball energy per unit charge is less than the mass
of the lightest Q-charged fermion) or slow (if the charge is
sufficiently large). The relic Q-balls can thus survive until present
and contribute to the dark matter in the universe.

Under very general conditions the scalar condensate can be described by
the classical equations of motion in the effective potential with
initial conditions determined by the high-scale physics or inflation.
It is convenient to write the complex field $\phi=R e^{i\Omega}$ in
terms of its radial component and a phase, both real functions of the
space-time coordinates. We are interested in the evolution of the
scalar field in the low-energy domain, where the baryon number
violating processes are suppressed, and we will assume that the scalar
potential preserves the U(1) symmetry: $U(\vf)=U(R)$, where $U(R)$ may
depend on time explicitly. The classical equations of motion in the
spherically symmetric metric $ds^2=dt^2-a(t)^2 dr^2$ are

\begin{eqnarray}
\ddot \Omega +3 H \dot \Omega - \frac{1}{a^2(t)} \Delta \Omega +\frac{2
\dot
  R}{R}
\dot \Omega - \frac{2}{a^2(t) R} (\dd_i \Omega)(\dd^i R) & = & 0,   
\label{eqnmtn1}
\\
\nonumber \\
\ddot R + 3 H\dot R - \frac{1}{a^2(t)} \Delta R - 
\dot \Omega^2 R + \frac{1}{a^2(t)} (\dd_i \Omega)^2
R + (\dd U/\dd R) & = & 0,
\label{eqnmtn2}
\end{eqnarray}
where dots denote the time derivatives, and the space coordinates are
labeled by the Latin indices that run from 1 to 3. The Hubble constant
$H=\dot a/a$, where $a(t)$ is the scale factor, is equal to $t^{2/3}$
or $t^{1/2}$ for the matter or radiation dominated universe,
respectively.

A spatially homogeneous solution of equations (\ref{eqnmtn1}) and
(\ref{eqnmtn2}) has vanishing gradients of $R$ and $\Omega$. If one
neglects the expansion of the universe ($H=0$) and if $\dot R=0$, the
solution describes a universe filled with Q-matter \cite{coleman}. If
the theory admits Q-balls, Q-matter can become unstable with respect to
its break-up into Q-balls. In a static universe, the conditions of
stability were analyzed in Ref. \cite{klee}. The results cannot be
applied directly to the case of the expanding universe, where the
equations of motion are time-dependent and the red shift of linear
perturbations is important. Under some conditions spatial
inhomogeneities may destabilize a given solution of equations
(\ref{eqnmtn1}) and (\ref{eqnmtn2}).

{}From the equations of motion (\ref{eqnmtn1}) and (\ref{eqnmtn2}), one
can derive the equations for small perturbations $\delta \Omega$ and $
\delta R$:

\begin{eqnarray}
\ddot{\delta \Omega} + 3 H \dot{(\delta \Omega)}
- \frac{1}{a^2(t)} \Delta (\delta \Omega) +\frac{2 \dot
  R}{R} \dot{(\delta \Omega)}+ \frac{2 \dot \Omega }{R}
\dot{(\delta R)}
- \frac{2\dot R \dot \Omega }{R^2} \delta R
 & = & 0, \label{eqndelta1} \\
\nonumber \\
\ddot{\delta R} + 3 H \dot{(\delta R)}
- \frac{1}{a^2(t)} \Delta (\delta R)
-2 R \dot \Omega \dot{(\delta \Omega)}+ U'' \delta R - \dot \Omega^2
\delta R & = & 0.
\label{eqndelta2}
\end{eqnarray}

In general, the analysis of instabilities must be done numerically.
However, one can identify the growing modes and estimate the length
scale of the resulting pattern by using a WKB approximation. This
approximation is valid as long as the background solution varies slowly
with time.

To examine the stability of a homogeneous solution $\vf(x,t)=\vf(t)
\equiv R(t) e^{i \Omega(t)}$, let us consider a perturbation $\delta R
, \delta \Omega \propto e^{S(t) - i \vec{k}\vec{x}} $ and look for
growing modes, ${\rm Re} \, \alpha>0$, where $\alpha = dS/dt $. The
value of $k$ is the spectral index in the comoving frame and is
red-shifted with respect to the physical wavenumber $\tilde{k}=k/a(t)$
in the expanding background. Of course, if the instability develops,
the linear approximation is no longer valid. However, we assume that
the wavelength of the fastest-growing mode sets the scale for the high
and low density domains that evolve into Q-ball eventually. This
assumption can be verified {\it post factum} by comparison with a
numerical solution of the corresponding partial differential equations
(\ref{eqnmtn1}) and (\ref{eqnmtn2}), where both large and small
perturbations are taken into account.

The dispersion relation follows from the equations of motion:

\beq
\left [
\alpha^2+ 3H\alpha +\frac{k^2}{a^2} + \frac{2 \dot R}{R} \alpha \right
]
\left [ \alpha^2+ 3H\alpha +\frac{k^2}{a^2} - \dot \Omega^2+U''(R)
\right ]+
4 \dot \Omega^2 \left [ \alpha - \frac{\dot R}{R} \right ] \alpha = 0.
\label{dr}
\eeq

If $(\dot \Omega^2-U''( R ))>0$, there is a band of growing modes that
lies between the two zeros of $\alpha(k)$, $0<k<k_{max}$, where

\beq
k_{max}(t)=
a(t) \sqrt{\dot \Omega^2-U''( R )}.
\label{band}
\eeq

If $k_{max}(t)$ defined by relation (\ref{band}) is constant or growing
with time, then each mode has an unlimited time to develop and the size
of a resulting inhomogeneity in the comoving frame is determined by the
value of $k$ that corresponds to the maximum of $\alpha$.

If, on the other hand, $k_{max}(t)$ decreases with time, then the modes
may not have the time to grow sufficiently before they are red-shifted
from the resonance. A constant or growing with time width of the band
(\ref{band}) is a sufficient (although not necessary) condition for the
instability.

Quantitatively, the  amplification of a given mode $k$
is characterized by the exponential of

\beq
S(k)=\int \alpha(k,t) dt.
\label{S}
\eeq
We look for the value of the best-amplified mode with wavenumber
$k_{best}$ that maximizes $S(k)$. The instability develops when
$e^{S(k_{best})}\gg 1$. The amount of amplification needed in a given
system is determined by the size of the fluctuations. These can be
thermal, or of a different origin (see, {\it e. g.}, the discussion in
Ref.\cite{pert}). The critical value of $S(k_{best}) \sim \ln (R/\delta
R)$ signals the non-linear regime and corresponds to the breakdown of
the homogeneous solution.

As an example, let us consider the following potential:

\beq
U(\vf) =  m_{_S}^4 \, \ln \left(1+ {\vf^\dagger \vf \over m_{_S}^2}
\right) -
c H^2 \vf^\dagger \vf +
\frac{\lambda^2}{m_{_P}^2} (\vf^\dagger \vf)^3,
\label{ptn}
\eeq
where $c$ and $\lambda$ are some constants and $H\sim 1/t$. Such
potential can arise in theories with supersymmetry breaking
communicated at the scale $m_{_S} \ll m_{_P}$ \cite{gia1,gmm} 
and is similar to those discussed in Ref. \cite{drt,gmm} in connection with
baryogenesis. The mass term proportional to the Hubble constant appears
naturally from the 
K\"ahler potential as $\int d^4 \theta (\langle \chi^\dagger \chi
\rangle/m_{_P}^2) \, \vf^\dagger \vf \ $ \cite{gia}, where $\chi$ is
some field with non-zero density, whose $F$-component has a VEV. This
term essentially characterizes supersymmetry breaking by the finite
energy density that dominates the universe at time $t$.

We will assume that the universe before Q-balls formation is matter
dominated. For large $|\vf|$, the logarithmic term in (\ref{ptn}) can
be neglected and the equations of motion admit an adiabatic solution
for $R(t) \sim 1/t^{1/2}$ and $\omega \equiv \dot \Omega \sim 1/t$,
that follows closely the location of the minimum of $U$. The
conservation of charge per comoving volume implies $\dot \Omega(t)
R^2(t) a(t)^3 = const$.

It is clear from equation (\ref{band}) that the band of resonance,
bounded from above by $k_{max}\sim t^{-1/3}$, becomes narrower with
time. In other words, each mode in the amplification region is
red-shifted faster than $k_{max}$, and the amplification may not be
sufficient. We note that the time dependence of $\omega$ is the same as
that of $H$, both are inversely proportional to time.

For small $|\vf| \sim m_{_S}$, the first term in equation (\ref{ptn})
dominates: $U(\vf)\approx m_{_S}^2 \vf^\dagger \vf$. In this case, the
equation of motion for $R$ is linear and its solution is $\vf \sim
1/t$, with a constant $\omega = m_{_S}$. Clearly, in this regime there
is an unlimited growth of all modes in the band of instability because
$k_{max} \sim a(t) $ is growing with time and the band widens.

\begin{figure}
\setlength{\epsfxsize}{3.3in}
\centerline{\epsfbox{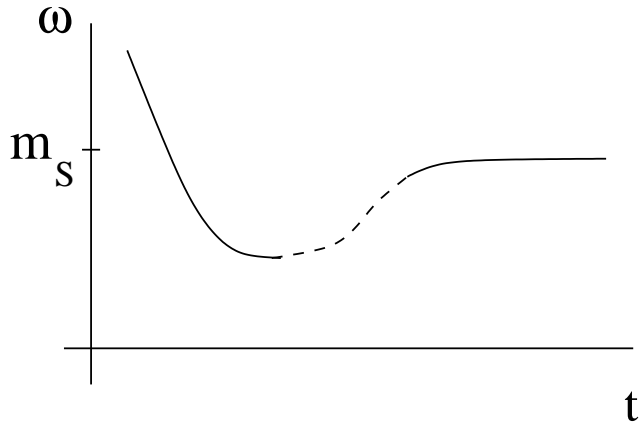}}
\caption{
At early times {$\omega$} changes at, roughly, the same rate as the
expansion of the universe. All modes of instability are red-shifted
before they can grow sufficiently. However, at later times {$\omega$}
approaches a constant value {$\sim m_{_S}$} and the instabilities
develop. The size of a domain is determined by the fastest-growing mode
at the time when the rate of change in {$\omega $} becomes different
from {$1/t$}.
}
\label{fig1}
\end{figure}

Qualitatively, the behavior of $\omega$ is shown in Fig. \ref{fig1}. As
long as the logarithmic term in the potential can be neglected,
$\omega$ changes as $1/t$, roughly, at the rate of the expansion of the
universe, and its value is of order $H$. However, at later times
$\omega$ approaches a plateau $\omega \sim m_{_S}$. A characteristic
length of growing instability is determined by the fastest-growing mode
at the time when the rate of change in {$\omega $} becomes different
from {$1/t$}.

To estimate a charge of Q-balls produced we assume that Q-matter was
formed at some time $t_0$ with charge density $q_0\equiv q(t_0) =
\omega_0 \vf_0^2$. In a matter dominated universe $q(t) = q_0
(t_0/t)^2$. Naturally, the wavelength $1/k \sim \chi/H $ of the
best-amplified mode cannot exceed the size of the horizon at the time
$t_i$ when the instability develops, $\chi < 1$. The total charge of
Q-matter inside the part of horison corresponding to the best amplified
model gives an estimate for the charge of Q-ball:
\beq
Q \sim q_0 t_0^2 t_i \chi^3 .
\label{size}
\eeq

\begin{figure}
\setlength{\epsfxsize}{3.3in}
\centerline{\epsfbox{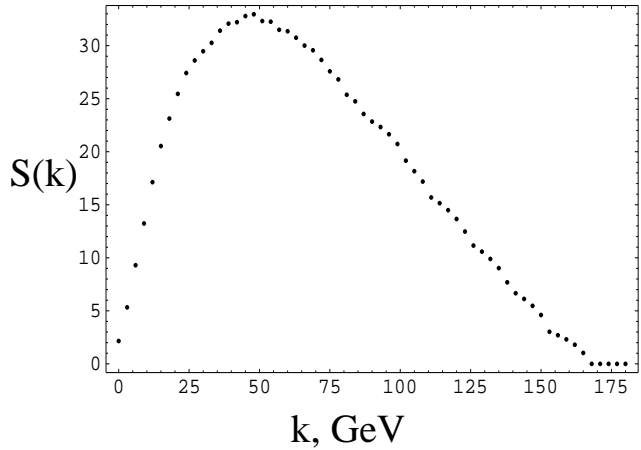}}
\caption{ 
The amplification {$S(k)= \int_{t_0}^{t_{max}} \alpha(k,t) dt$} of
growing modes is computed numerically for different values of {$k$}.
The fastest-growing mode has the wave number {$k \sim 10^2 H $} at the
time when the instability develops. An individual Q-ball gathers charge
from, roughly, {$10^{-2}$} of the size of the horizon.
}
\label{fig_S}
\end{figure}

\begin{figure}
\setlength{\epsfxsize}{3.3in}
\centerline{\epsfbox{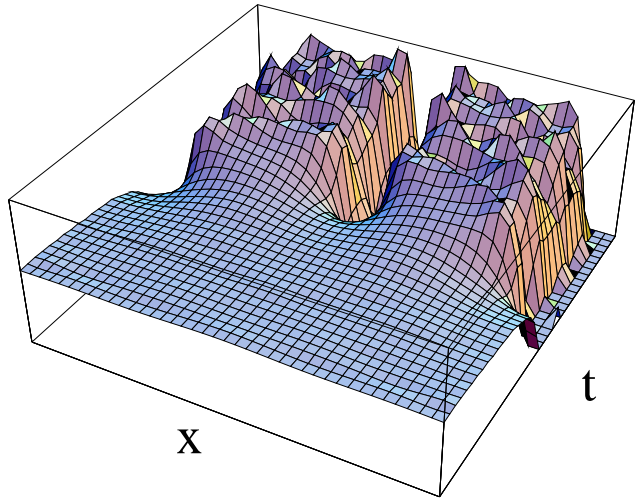}}
\caption{ 
The charge density per comoving volume in (1+1) dimensions for a sample
potential analyzed numerically during the period when the spatially
homogeneous condensate breaks up into high- and low-density domains.
Two domains with high charge density are expected to form Q-balls.
}
\label{fig_charge}
\end{figure}

The time $t_i$ and parameter $\chi$ are to be found from numerical
calculations. We performed them for a potential (\ref{ptn}) for
different sets of parameters. The amplification $S(k)$ of a given mode
was found through numerical integration. We took the initial time of
the Q-matter formation to be $t_0=10^{-2}$ GeV$^{-1}$. For $\lambda  
=1/2$,
$c=1$, $m_{_S}=10^4$ GeV and $q_0 = 10^{24}$GeV$^3$ (this corresponds
to $\vf_0= 2.5\times 10^{10}$ GeV and $\omega_0= 1.6\times 10^3$ GeV)
we observed that the best-amplified mode was two orders of magnitude
smaller than the horizon size, $\chi \simeq 10^{-2}$, at the time when
the instability developed ($S_{max}(k) \simeq 30$ at $t_i \simeq 150$
GeV$^{-1}$). This corresponds to the average charge of emerging Q-balls
of order $10^{16}$. For a different set of parameters, $m_{_S}= 10^2$
GeV,  $\lambda = 0.5 \times 10^{-3} $ and $q_0=10^{25}$ GeV$^3$
(which correspond to $\vf_0= 2.8\times 10^{11}$ GeV and $\omega_0=
1.3\times 10^2$ GeV), we obtain $t_i \simeq 1.5\times 10^5$ GeV$^{-1}$,
$\chi \simeq 10^{-2}$ and $Q \sim 10^{20}$. The band of unstable modes
is shown in Figure~\ref{fig_S}. Of course, the size of Q-balls depends
on a model. We have demonstrated, however, that very large Q-balls can
be born from an initially homogeneous scalar condensate. In particular,
the flatter, than logarithmic, potentials produce even larger Q-balls.
It would be desirable to study the instabilities beyond the linear
analyses performed so far, when the solution is significantly different
from the original homogeneous one, as illustrated in Figure
\ref{fig_charge}. We leave this for future work.

\section{Surviving Q-balls as dark matter}

It is beyond the scope of this paper to examine in detail the
implications of solitonic dark matter formed from the breakdown of a
scalar condensate. We will limit our discussion to a few general
remarks.

We have seen that B-balls formed through the instability can easily
satisfy the bound (\ref{stableQ}). If finite temperature effective
action admits B-balls, they will survive till present. If not, they may
or may not be eroded by the thermal plasma, depending on the model
parameters and temperature $T_i$ at the time of the Q-ball formation.  
From
equation (\ref{erosion}) they are not destroyed if $T_i < m_\vf(m_\vf
Q/M_0)^{\frac{1}{2}}$. Surviving B-balls would contribute to the dark
matter in the Universe. The amount of the dark matter is determined by
the efficiency of the conversion process that leads from a uniform
condensate to B-balls. As the instability develops, part of the charge
(baryon number) stored in the condensate can be lost in decay and give
rise to the baryonic matter. It seems reasonable to assume that the
amount of baryonic charge locked inside the Q-balls can be roughly
comparable (within a few orders of magnitude) to the amount of charge
that leaked out with individual nucleons.

If the scalar condensate that evolves into Q-balls has a lepton charge,
but no baryon charge, then a Q-ball can evaporate \cite{evap} by
emitting lepton number from the surface with (nearly massless)
neutrinos. We have seen that, for the lifetime of an L-ball to exceed
the age of the Universe, its charge should be $Q>10^{32}$ according to
equation (\ref{decay}). Smaller L-balls must have evaporated by now.
Decay of the L-balls during or after the nucleosynthesis may have
important ramifications as it would lead to the entropy increase and,
if it happens at later times, could distort the spectrum of the cosmic
microwave background radiation. This may lead to constraints on the
parameters of the MSSM, as well as on the initial conditions after
inflation.

Our discussion has bearing on the issue of the so called moduli problem
\cite{moduli,gia}. Small couplings to matter fields are typical for the
moduli fields of string theory. This poses a serious problem for
cosmology since the energy density carried by the coherent motion of
the moduli condensates does not decay until after the nucleosynthesis.
The products of such decay can over-produce the entropy, as well as
destroy the nuclei. The mechanism we have described may help alleviate
the moduli problem by incarcerating moduli inside slowly evaporating
Q-balls. On the one hand, this slows down the decay, thus allowing for
the moduli Q-balls to still be present in the Universe and contribute
to dark matter. On the other hand, whatever decay would have taken
place, the detrimental effect of the energetic photons on the nuclei is
confined to the close vicinity of Q-balls separated by some vast
unaffected areas, where the standard nucleosynthesis can take place.

Let us suppose that in the present Universe there is, roughly, the same
(to within a few orders of magnitude) baryon number in baryonic Q-balls
and in ordinary matter. According to our estimates, the charge of an
individual stable B-ball may be from $10^8$ to $10^{20}$ or more; and
the corresponding number density of Q-balls is then $10^{-20}$ to
$10^{-8}$ per nucleon. Perhaps, even much larger Q-balls may emerge
from the scalar condensate in models different from ours. The larger
Q-balls from this range may be difficult to detect because they are
extremely rare. However, the possibility of smaller, and, hence, more
populous B-balls as a dark matter candidate seems to be very appealing
from an experimental point of view. Some of the heavy relic Q-balls,
attracted by the gravitational fields of stars and planets, can sink to
the center and remain there. It is conceivable that the deep interior
of the small planets might become accessible for exploration in the
future and reveal storages of primordial Q-balls. Other solitons may be
present in the interstellar medium. However, since their number density
is expected to be very small, detection of such objects seems to be a
great challenge to the experimentalists' ingenuity. It is also of
interest to understand the effect of large Q-balls on the structure
formation in the early universe.

We thank G.~Dvali, H.~Murayama and P.~Tinyakov for helpful discussions.

\end{document}